# The inherently absent 2-dimensional electron gas in ultra-pure GaN/AlGaN heterostructures

S. Schmult[1*], S. Wirth[2], V.V. Solovyev[3], R. Hentschel[4], A. Wachowiak[4], A. Großer[4], I.V. Kukushkin[3,5], and T. Mikolajick[1,4]

[1] TU Dresden, Electrical and Computer Engineering, Institute of Semiconductors and Microsystems, Nöthnitzer Str. 64, 01187 Dresden, Germany

[2] Max-Planck-Institute for Chemical Physics of Solids, Nöthnitzer Str. 40, 01187 Dresden, Germany

[3] Institute of Solid State Physics RAS, 142432 Chernogolovka, Moscow district, Russia

[4] Namlab gGmbH, Nöthnitzer Str. 64, 01187 Dresden, Germany

[5] National Research University Higher School of Economics, 101000 Moscow, Russia

* Corresponding author: stefan.schmult@tu-dresden.de

**Abstract**

Gallium nitride (GaN) has emerged as an essential semiconductor material for energy-efficient lighting and electronic applications owing to its large direct bandgap of 3.4 eV. Present GaN/AlGaN heterostructures seemingly feature an inherently existing highly-mobile 2-dimensional electron gas (2DEG), which results in normally-on transistor characteristics. Here we report on an ultra-pure GaN/AlGaN layer stack grown by molecular beam epitaxy, in which such a 2DEG is absent at 300 K in the dark, a property previously not demonstrated. Illumination with ultra-violet light however generates a 2DEG at the GaN/AlGaN interface and the heterostructure becomes electrically conductive. At temperatures below 150 K this photo-conductivity is persistent with an insignificant dependence of the 2D channel density on the optical excitation power. Residual donor impurity concentrations below $10^{17}$ cm$^{-3}$ in the GaN/AlGaN layer stack are one necessity for our observations. Fabricated transistors manifest that these characteristics enable a future generation of normally-off as well as light-sensitive GaN-based device concepts.

**Main**

Group-III-Nitrides and their heterostructures have become key elements in energy-saving daily life applications, such as white solid-state lighting [1] and short wavelength lasers for data storage concepts [2]. Single-interface gallium nitride/aluminum gallium nitride (GaN/AlGaN) heterostructures feature a highly-mobile electron channel with mobilities exceeding 2000 cm$^2$/Vs at room temperature [3]. Combined with the large bandgap of 3.4 eV and the associated high critical breakdown field strength as well as thermal and chemical stability of this compound material, GaN-based devices increasingly emerge in power and high-voltage applications [4,5]. On the other hand, GaN possesses intrinsic material properties like a large effective electron mass and a large electron g-factor, which allows for exploiting an extended regime of mesoscopic physics compared to its compound semiconductor counterpart gallium arsenide (GaAs) [6-8].

The aforementioned mobile channel at the GaN/AlGaN interface is a 2-dimensional electron gas (2DEG) in close proximity to the surface (10 – 50 nm) with record low temperature electron mobility in excess of 100000 cm$^2$/Vs [9]. Due to polarization changes at the GaN/AlGaN interface a confinement potential is formed [10] which is usually flooded with electrons (Fig. 1). One possible resource of these charges are surface states, from which the electrons are transferred into the confinement potential [11]. Following this view, in such structures the generation of the 2DEG does not require intentionally introduced donors which would also act as efficient Coulomb scattering centers. However, one disadvantageous consequence of this undeliberately existing 2DEG for use in lateral heterostructure field-effect transistors (HFETs) is the associated normally-on switching characteristics, a drawback for fail-safe applications. Technologically this normally-on characteristics can be converted into the desired normally-off fashion, for example by an additional recess etch step, an extra p-type GaN or AlGaN layer beneath the gate electrode [4,5] or a hybrid GaN/Si cascode technology approach [12], but all of which require additional processing efforts and costs.

When inspecting the band diagram of the underlying GaN/AlGaN heterostructure in more detail, impurities or defects in the GaN or AlGaN layers represent another possible source of the channel charges, as they will be transferred into the inherently existing polarization-induced confinement potential to minimize their potential energy (Fig. 1). To exclude the GaN or AlGaN layers as a possible source of channel charges one must be able to synthesize ultra-pure material with an ionized impurity background significantly below $10^{17}$ cm$^{-3}$, as will be discussed later. Here we demonstrate by using an ultra-pure GaN/AlGaN layer stack that the predominant source of the charges forming the 2DEG is represented by bulk impurities in the GaN/AlGaN layer stack rather than by surface states.

Results:

Under investigation is a GaN/AlGaN heterostructure with a low level of residual impurities (approximately $2 \cdot 10^{16}$ cm$^{-3}$). The GaN/Al$_{0.06}$Ga$_{0.94}$N/GaN layer stack with respective thicknesses of 1 µm/16 nm/3 nm (inset Fig. 2) is grown by molecular beam epitaxy (MBE) on insulating hexagonal GaN. The only detectable impurity in the unintentionally doped (uid) MBE stack is oxygen, and its level of incorporation into GaN and AlGaN is defined by the sample temperature during epitaxial growth [13,14]. Comparable low background impurity levels are also found in MBE-grown GaN material from other laboratories [15].

The discussed GaN/AlGaN heterostructure with ohmic contacts shows electrically insulating behavior in the dark at low temperatures as well as at 300 K, as discussed later. Under ultra-violet (UV) illumination the sample turns conductive, which is explained by the formation of a 2DEG at the GaN/AlGaN interface. In addition, below 150 K the conductivity becomes persistent in the dark, i.e. several hours after blanking the UV light the electron density reaches an equilibrium (persistence) value independent of the initial excitation power of ~ 2 · $10^{12}$ cm$^{-2}$ (Fig. 2), a density which is expected and demonstrated for this stack architecture [16] and previously repeatedly measured [13,14]. When illumination is repeated, the density rises instantly to a value depending on the excitation power. When UV light is blanked again, the density decreases abruptly, followed by a slow relaxation prior to reaching the equilibrium value again only after hours.

Albeit the equilibrium channel density is power-independent, the value under steady UV illumination depends on the excitation power (Fig. 3). Magneto-transport measurements at 0.5 K under steady illumination yield Shubnikov-de Haas oscillations with clear zero longitudinal resistance values around 10 T. These variations in the longitudinal conductivity not only confirm the 2D nature of the generated conductive channel, but also demonstrate the absence of other (parasitic) current paths except this 2D channel. Only for large excitation power the transport traces are deteriorated and clear zero longitudinal magneto-resistance is absent for B < 15 T, which is explained by sample heating to well above 0.5 K (trace under 4 mW excitation in Fig. 3).

The observed Shubnikov-de Haas oscillations also allow for the assignment of Landau level filling factors and the subsequent extraction of the 2D channel density. In the described experiment a change in steady UV excitation power over almost 5 orders in magnitude only accounts for a ~ 50 % increase of the 2D channel density up to 3.0 · $10^{12}$ cm$^{-2}$, compared to its equilibrium value of ~ 2 · $10^{12}$ cm$^{-2}$ (see values given in Fig. 3).

The insulating behavior at 300 K in the dark combined with the possibility to generate a conductive channel under UV illumination will facilitate new routes for device application by exploiting the electrostatic population of the confinement potential at the GaN/AlGaN interface. This is equivalent to the technologically desired normally-off switching characteristics of lateral HFETs. A normally-off device processing scheme includes an overlap of the gate layer with the ohmic source and drain contacts. To achieve this, lithographically defined lateral HFET test structures were covered with a highly insulating aluminum oxide layer subsequent to ohmic contact formation (Fig. 4 (a) and (b)). This architecture allows for an electrostatic response of the entire region beneath the gate electrode, including the proximity of source and drain contacts.

Lateral HFETs based on our GaN/AlGaN stack were processed with the described normally-off scheme. As shown in Fig. 4 (d), they operate in enhancement mode (normally-off) at 300 K in the dark, such as their GaAs–based counterparts [17] and inversion channel metal-oxide-semiconductor (MOS) FETs based on silicon [18]. Under steady UV illumination normally-on behavior is restored, and fades away after blanking the light. In both cases the conducting 2D channel is located at the position of the confinement potential at the GaN/AlGaN interface (Fig. 5), as verified in capacitance vs. voltage measurements. The switching characteristics of our FETs establish a trend set by the unintentionally incorporated oxygen: For high concentrations (~$10^{18}$ cm$^{-3}$) in the MBE-grown GaN material FETs cannot be turned off due to parasitic current paths, at medium values (~$10^{17}$ cm$^{-3}$) they operate normally-on [13,14] and ultimately, at a low level as realized here, normally-off.

Discussion:

The presented data strongly indicate that the confinement potential at the GaN/AlGaN interface is inherently not populated with charge carriers. Upon illumination with photon energies in the region of the GaN bandgap the channel is filled and a 2DEG - and only a 2DEG - as a conduction path is generated. This 2DEG persists at low temperature (i.e. after blanking the UV illumination) and exhibits an equilibrium electron density of ~$2 \cdot 10^{12}$ cm$^{-2}$. Further steady optical excitation will certainly add photo-generated electrons, though this only insignificantly impacts the channel density. The layer stack architecture and the resulting potential landscape set a limit for the number of electrons accommodating the 2DEG. This channel density robustness is demonstrated under steady flooding with photo-induced electrons over several magnitudes in excitation power. The low impurity level of ~$2 \cdot 10^{16}$ cm$^{-3}$ (with an even lower value for ionized impurities below $10^{16}$cm$^{-3}$) in the epitaxial GaN/AlGaN stack is the hitherto unidentified source of the channel charges. A back-on-the-envelope calculation yields a number of $10^{13}$ cm$^{-2}$ electrons, when $10^{17}$ cm$^{-3}$ ionized impurities are distributed over the 1 µm thick MBE stack. In this case a 2DEG is present independent of illumination.

The transistor transfer characteristics $I_{DS}(V_{GS})$ at 300 K at low drain-source voltage ($V_{DS}$) with and without UV illumination support this interpretation (Fig. 4 (d)). In the dark, the HFET exhibits transfer characteristics typical for normally-off devices with a threshold voltage ($V_{th}$) around +1.8 V. With increasing gate bias the drain current first remains constant at the measurement resolution limit, and then increases exponentially in the subthreshold region towards the threshold voltage with a subthreshold swing of about 80 mV/decade, a value close to the thermal limit of about 60 mV/decade. Assuming a similar gate coupling in the entire subthreshold region, one can estimate from the ratio between the ideal subthreshold swing and the measured one that the confinement potential at the GaN/AlGaN interface is about 1.4 eV above the Fermi-energy at the potential of the source region for $V_{GS}$ = 0 V. When $V_{GS}$ is raised above $V_{th}$ the drain current linearly increases with $V_{GS}$ since the confinement potential is lowered below the potential of the source region and the 2DEG charge density is linearly tuned with $V_{GS}$ - $V_{th}$ given by the capacitive coupling of the series capacitances from gate dielectric, GaN cap and AlGaN barrier layer (Fig. 4 (c)). The drain current tends to saturate upon filling the confinement potential with a maximum level of charge carrier density, so electrons can easily overcome the AlGaN barrier towards the GaN cap / dielectric interface.

Those transferred electrons could be captured in traps at the GaN cap / dielectric interface as observed in hysteresis phenomena with a shift towards positive $V_{GS}$ of transfer curves measured during back-sweeping to lower $V_{GS}$ again. However, negligible hysteresis (< 0.1 V) was observed in double sweep measurements of transfer characteristics with highest $V_{GS}$ values below overloading the 2DEG density. Moreover, once electron trapping occurred, the transfer characteristics could be shifted back to the original position on the $V_{GS}$ scale by stressing the device at high negative $V_{GS}$ values, e.g. -8 V, for several minutes for complete de-trapping. Additionally, double-pulsed measurements of transfer characteristics with a repetition rate of 20 Hz and pulse width 1 ms performed at quiescent gate-source voltages of -3 V, 0 V and 3 V always result in positive threshold voltages > 1.5 V. Therefore, we can exclude trapped electrons in the dielectric or at the GaN cap / dielectric interface as the root cause of the positive threshold voltage in the dark case.

The same scenario holds for photo-excited electrons. Once the channel is filled up to its equilibrium density, additional electrons will favorably overcome the AlGaN barrier and transfer into the GaN cap / surface by tunneling or thermal stimulation. This very likely explains the density robustness of the channel demonstrated in Fig. 3.

**Methods**

The discussed heterostructure was grown along the Ga-polar c-axis in a VG80H MBE system at a temperature slightly below the rapid Ga desorption point on an insulating 2 inch GaN bulk substrate prepared by ammonothermal synthesis with a density of threading and mixed dislocations < $10^5$ cm$^{-2}$. Ga and Al with nominal purity of 8N and 6N5, respectively, were evaporated from filament effusion sources, active Nitrogen (nominal purity 10N) was supplied by a radio-frequency plasma source. The growth rate for GaN amounts to 240 nm/h. After decollating pieces of the 2 inch wafer, Hall bars and transistor test structures were lithographically defined and mesa structures patterned in a reactive ion etch step using a chlorine-based plasma. Ti/Al/Ni/Au stacks annealed at 800 °C in nitrogen atmosphere for 30 s serve as ohmic contacts. For transistors the entire surface (including contacts, mesa and etched regions) was covered with 27 nm $Al_2O_3$ by atomic layer deposition. A 100 nm thick Ti/Au film serves as gate electrode.

Background impurity levels have been verified by secondary ion mass spectroscopy (SIMS), demonstrating an unintentional oxygen level of ~ $2 \cdot 10^{16}$ cm$^{-3}$ in the grown GaN material.

Temperature- and illumination-dependent magneto-transport measurements between 2 K and 300 K were performed in a Physical Property Measurement System (PPMS) from Quantum Design with optical access. A low-frequency (17 Hz) lock-in technique with excitation current of 100 nA was employed.

A $^3$He cryostat equipped with a superconducting magnet (up to 15 Tesla) was used for low temperature magneto-transport studies. The sample photoexcitation at 325 nm wavelength was realized with the fiber-coupled output of a He-Cd laser. The large distance of ~ 20 mm between the bare fiber end (numeric aperture = 0.22) and the sample surface enabled a good uniformity of both the illumination and the resulting 2DEG density as confirmed by the onset of SdH oscillations. The photoexcitation power was attenuated by optical filters and then measured at the fiber input, the fiber optical transmission amounts to ~ 80 %. Magneto-transport data were recorded using a low-frequency lock-in technique.

The capacitance vs. voltage C($V_{GS}$) measurements were carried out by sweeping a quasi-static DC bias modulated with an AC signal with a frequency of 20 kHz and with an amplitude of 30 mV using an Agilent B1505A Power Device Analyzer equipped with a capacitance measurement unit. The DC bias voltage at the gate electrode was swept in the direction from lower to higher voltages with a staircase-like sweep pattern and a rate of approximately 0.5 V/s. Depth profiles of the charge carrier density were extracted by applying the data transformation according to reference [10] after subtracting a constant offset capacitance, which accounts for the direct overlap capacitance described in the text. Transfer $I_D(V_{GS})$ characteristics are obtained with the same device analyzer. Double-pulsed transfer characteristics measurements were performed with the same device analyzer at a repetition rate of 20 Hz and pulse width of 1 ms at quiescent gate-source voltage values between -3 and 3 V applied for 49 ms.

**Acknowledgements**

We acknowledge support in device processing from K. Nieweglowsky, J. Gärtner and N. Szabo and appreciate constructive discussions with A. Ruf. The RAS part of the work was financially supported by the Russian Foundation for Basic Research. The NaMLab gGmbH portion of the work was financially supported by the German Federal Ministry for Economic Affairs and Energy - BMWi (project no.: 03ET1398B).

**Author contributions**

S.S. and A.G. executed sample growth. S.S., A.G., V.V.S. and R.H. processed devices. S.S. and S.W. designed and performed PPMS magneto-transport experiments, analyzed and processed data. V.V.S. and I.V.K. designed and performed $^3$He magneto-transport experiments and analyzed and processed data. A.W. and R.H. designed and performed electrical transistor characterization, analyzed and processed data. S.S., S.W. and A.W. prepared the manuscript. T.M. designed experiments, analyzed data and participated in manuscript preparation. All the authors have read the manuscript and agree with its content.

**Figures**

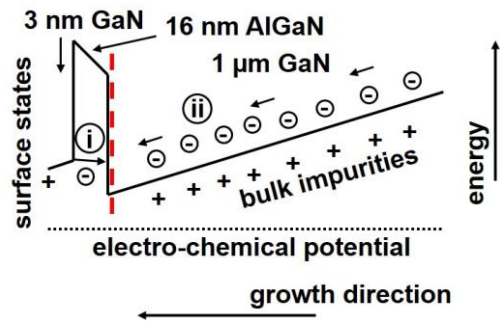

Fig. 1: Schematic conduction band diagram of the discussed GaN/AlGaN heterostructure (not to scale). Due to polarization discontinuity a triangular confinement potential is formed at the GaN/AlGaN interface (marked by the vertical dashed line). Two scenarios for filling this potential with electrons, and thus forming a 2DEG, are shown. In scenario i, electrons originate from surface states and tunnel through the AlGaN barrier to fill the confinement potential. In scenario ii, electrons resulting from background donor impurities minimize their potential energy by moving towards the GaN/AlGaN interface and create the 2DEG at the same. In order to rule out the latter case the level of bulk impurities must be very low (< $10^{17}$ cm$^{-3}$).

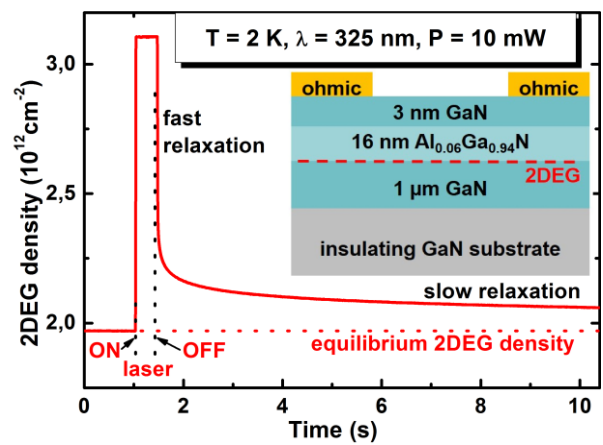

Fig. 2: Persistent equilibrium low temperature channel density of ~$2 \cdot 10^{12}$ cm$^{-2}$ (0 – 1 s) several hours after excitation with a UV laser at 10 mW of the discussed GaN/AlGaN heterostructure, in cross-section schematically displayed in the inset (not to scale). Further laser illumination (at t = 1 s) results in an instant rise of the channel density to a constant value of ~$3.1 \cdot 10^{12}$ cm$^{-2}$. After blanking the UV light the channel density shows initially a fast, followed by a slow relaxation trend, which restores after several hours the equilibrium value. The density is extracted from the time-dependent Hall voltage at 0.5 T in a laterally defined Hall bar. At 300 K the voltage-current characteristics shows ohmic behavior under steady UV illumination, as described in detail in the appendix, Fig. 6.

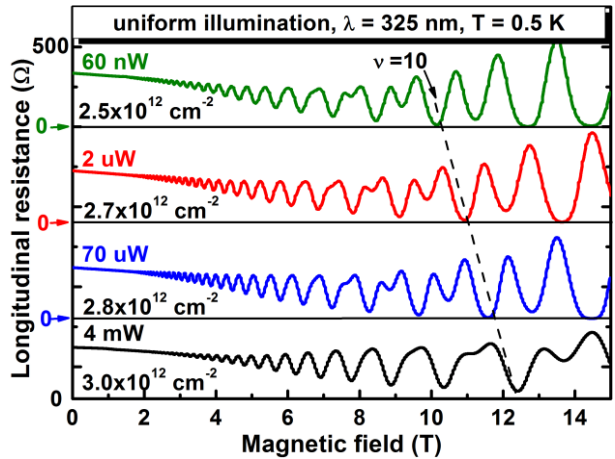

Fig. 3: Low temperature longitudinal magneto-resistance of a 400 µm wide Hall bar under steady photo-excitation with UV light at different excitation power. The observed Shubnikov-de Haas-oscillations and Zeroes in the longitudinal resistance clearly point at the 2D channel character with no parasitic current paths. The identification of Landau level filling factors (e.g. $\nu = 10$, dashed line) allows for the extraction of the respective electron densities. An increase in the illumination power of almost 5 orders in magnitude only raises the channel density by 20 %. Only at the largest excitation power the sample is overheated and its temperature raised above 0.5 K, reflected by the non-vanishing resistance values of the minima.

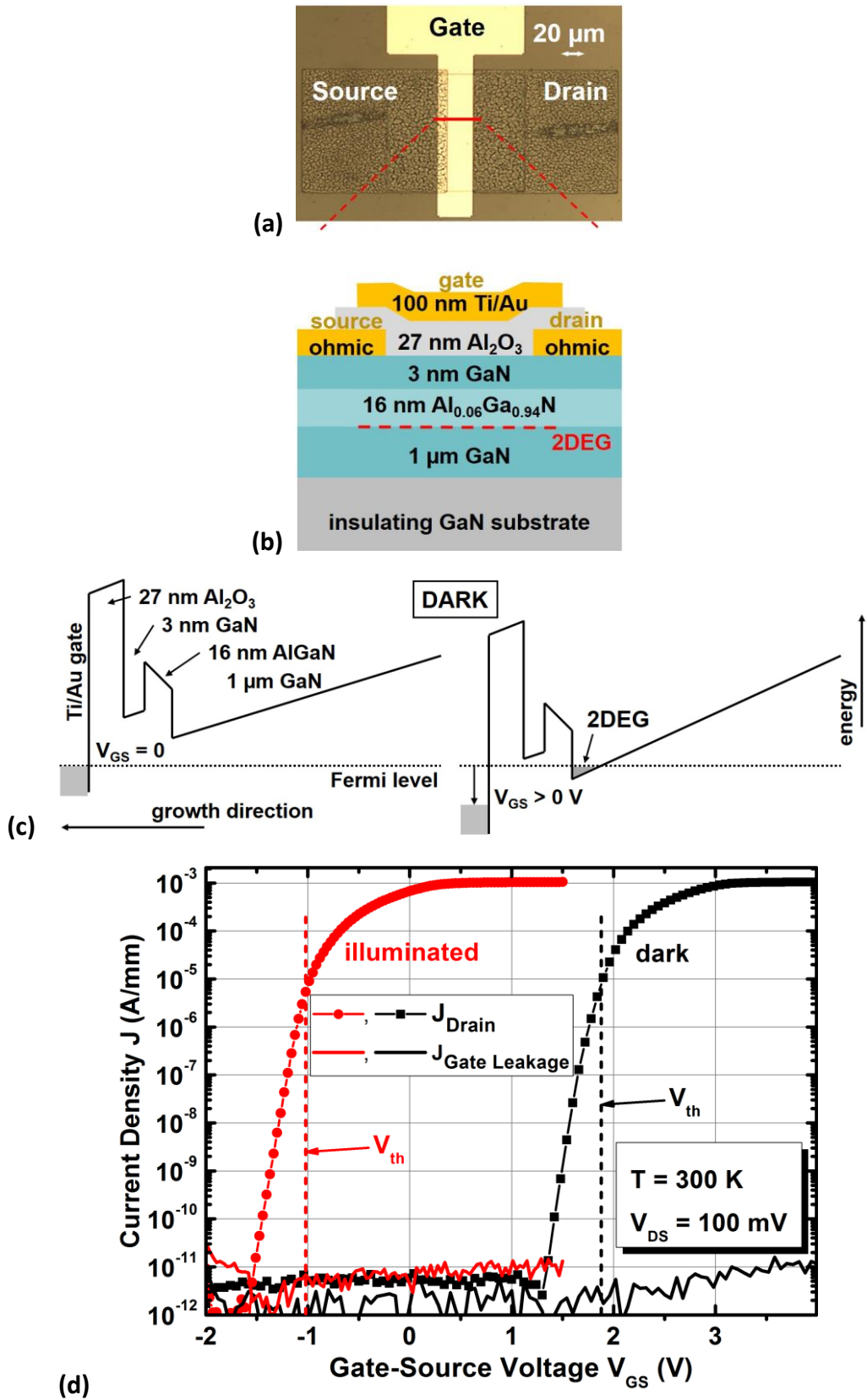

Fig. 4: Top-view photograph (a) and device schematics (b) illustrating the overlap of the gate contact with source and drain (clearly visible for the source contact, barely for the drain in (a)). This overlap is essential since the 2DEG is only electrostatically generated below the gate electrode in the dark. At gate-source voltage $V_{GS}$ = 0 V the confinement potential in the conduction band is energetically located above the Fermi level ((c) left, not to scale). By applying positive $V_{GS}$ the surface potential is lowered and finally the confinement potential can be populated by electrons originating from the source contact ((c) right). (d) Transfer characteristics of the drain current density scaled to gate width measured for a $V_{GS}$ bias sweep from low values in positive direction on a device with gate length of 15 µm and gate width of 100 µm in the dark and under UV illumination. Normally-off operation at 300 K in the dark is confirmed by the onset of the source-drain current $I_{Drain}$ (solid squares) at positive $V_{GS}$ ~ 1.3 V. In contrast, under UV illumination normally-on behavior is restored (solid circles), expressed by a source-drain current onset at negative $V_{GS}$. In both cases the on/off current ratio exceeds $10^8$ for a drain-source voltage $V_{DS}$ of 100 mV. The threshold voltages $V_{th}$ under UV illumination and in the dark are marked by vertical dashed lines. Gate leakage currents (solid lines) for both measurements are within the resolution limit, pointing at excellent insulation properties of the gate dielectric.

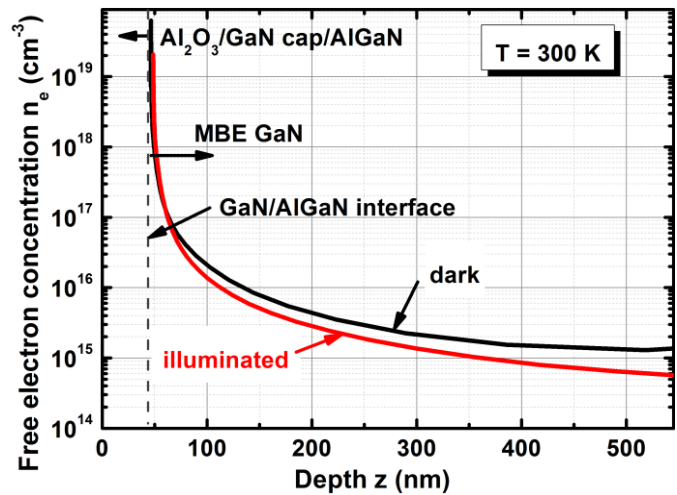

Fig. 5: Charge carrier profile of a $Al_2O_3$/GaN/AlGaN/GaN transistor layer stack (27 nm/3 nm/16 nm/1 µm) calculated from capacitance vs. voltage $C(V_{GS})$ data for both operation modes as shown in Fig. 4. The $Al_2O_3$/gate electrode (Ti/Au) interface is placed at 0 nm. The 2DEG for both cases is located at the GaN/AlGaN interface 46 nm beneath the gate electrode. Illumination is realized with a fluorescent lamp featuring emission lines in the UV spectral region.

# Appendix

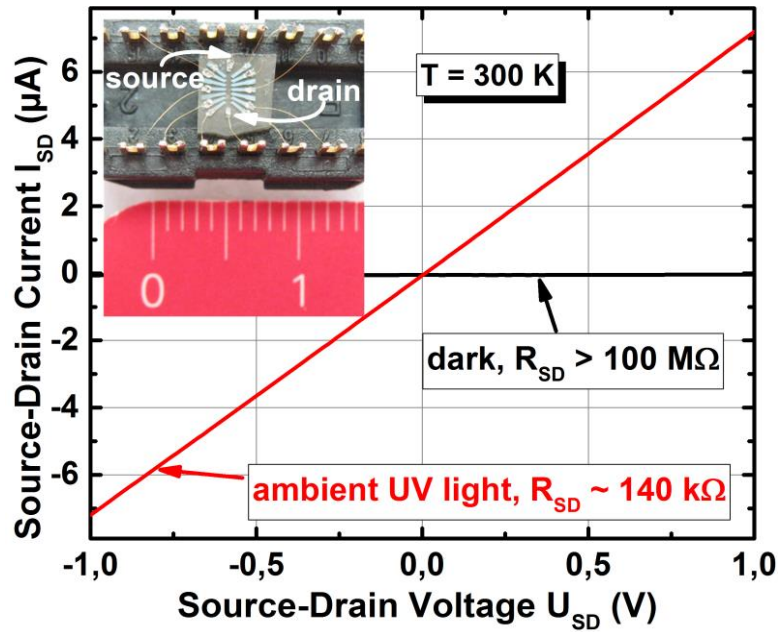

Fig. 6: Source-drain current vs. voltage characteristics in the dark and under ambient UV illumination of the processed and wired Hall bar without gate dielectric layer as sketched in Fig. 2. In the dark the resistance exceeds 100 MΩ, whereas under illumination the value decreases by almost 3 orders of magnitude and demonstrates ohmic behavior. The inset shows a photograph of the Hall bar in comparison with a centimeter ruler with the source and drain contacts indicated. Ohmic contacts around the Hall bar and their leads are visible, the lateral mesa definition itself is not seen due to the transparent character of the material. The Hall bar is 100 µm wide and 2.4 mm long, the 14 symmetrically distributed voltage probes on both sides are separated by 300 µm.

Further details on lateral HFETs and the comparison under UV illumination and in the dark:

Under UV light illumination, the transfer curve of the mentioned FET shows the behavior of a normally-on device with threshold voltage in negative gate bias region around -1 V. Remarkably, the entire course of the transfer curve measured in the dark is nearly an exact copy, just shifted into the positive $V_{GS}$ direction (e.g. having the same subthreshold slope). Such a shift would be expected in case of a net positive fixed charge density with respect to the dark case either in the region from the confinement potential towards the gate electrode or in the direction of the GaN channel region towards the substrate. Such a charge density would cause an internal bias field that further energetically lowers the GaN/AlGaN confinement potential compared to the case in the dark at the same external $V_{GS}$ bias.

Further evidence for the 2DEG at the GaN/AlGaN interface as the current carrying channel in transistor on-state operation for both cases with and without light illumination is obtained from capacitance-voltage measurements between gate electrode and ohmic source/drain contacts of the transistor structure. After subtracting from the measured capacitance trace a constant capacitance value resulting from direct overlap of the gate electrode on the source/drain, charge carrier profiles as a function of the depth from the gate electrode were extracted (Fig. 5). In the dark and under steady UV-illumination, the peak maximum in charge carrier density at depth of 45 nm beneath the gate electrode reveals the 2DEG position. With increasing depth the charge carrier density quickly falls below $1 \cdot 10^{16}$ cm$^{-3}$ in the deeper MBE-grown GaN channel layer supporting the low background impurity concentration. C(V$_{GS}$) measurements done in this device configuration always seem to yield concentrations of free carriers in the thick GaN buffer layer below $10^{15}$ cm$^{-3}$, independent of the background donor impurity concentration in the range $10^{16}$ cm$^{-3}$ to $10^{17}$ cm$^{-3}$. Pointing at [13,14], the in such manner extracted low free-carrier concentration is a necessary, but not sufficient proof for ultra-pure MBE GaN material with an active impurity background of <$10^{16}$ cm$^{-3}$, as commonly stated [19].